\documentclass[reprint,amsmath,amssymb,aps,superscriptaddress,nofootinbib,twocolumn,10pt]{revtex4-1}
\usepackage[colorlinks,linkcolor=blue,citecolor=blue,urlcolor=blue]{hyperref}
\usepackage{graphicx,epstopdf}
\usepackage{dcolumn}
\usepackage{subfigure}
\usepackage{bm}
\usepackage{amsmath,amsfonts,amssymb,graphicx,multirow,dcolumn,bm,latexsym,soul,nicefrac}
\usepackage{acronym}
\usepackage{enumitem}
\usepackage{array}
\usepackage{multirow}
\usepackage{appendix}
\usepackage{longtable}
\usepackage{footnote}
\usepackage{amssymb}

\newcommand{\TRC}{MOE Key Laboratory of TianQin Mission, TianQin Research Center for Gravitational Physics \& School of Physics and Astronomy, Frontiers Science Center for TianQin, Gravitational Wave Research Center of CNSA, Sun Yat-sen University (Zhuhai Campus), Zhuhai 519082, China.}

\begin{document}

\title{Distinguish the environmental effects and modified theory of gravity with multiple massive black-hole binaries}

\author{Xulong Yuan}
\affiliation{\TRC}
\author{Jian-dong Zhang}
\email{zhangjd9@mail.sysu.edu.cn}
\affiliation{\TRC}
\author{Jianwei Mei}
\affiliation{\TRC}
\newacro{GR}{general relativity}
\newacro{GW}{gravitational wave}
\newacro{MBH}{massive black hole}
\newacro{MBHB}{massive black hole binary}
\newacro{EMRI}{extreme mass-ratio inspiral}
\newacro{IMRI}{intermediate mass-ratio inspiral}
\newacro{ppE}{parameterized post-Einsteinian}
\newacro{BH}{black hole}
\newacro{NS}{neutron star}
\newacro{BS}{boson star}
\newacro{ECO}{exotic compact object}
\newacro{LVK}{LIGO-Virgo-KAGRA}
\newacro{BBH}{binary black hole}
\newacro{BNS}{binary neutron star}
\newacro{NSBH}{neutron star-black hole}
\newacro{SIQM}{spin induced quadrupole moment}
\newacro{ASD}{amplitude spectral density}
\newacro{PSD}{power spectral density}
\newacro{SNR}{signal-to-noise ratio}
\newacro{PE}{parameter estimation}
\newacro{PN}{post-Newtonian}
\newacro{FIM}{fisher information matrix}
\newacro{TDI}{time delay interferometry}
\newacro{ISCO}{innermost stable circular orbit}
\newacro{PDF}{probability distribution function}
\newacro{EPS}{extended Press and Schechter}
\newacro{LISA}{Laser Interferometer Space Antenna}
\newacro{ppE}{parameterized post-Einstein}
\newacro{DF}{dynamical friction}
\newacro{DM}{dark matter}
\newacro{IMRI}{intermediate mass-ratio inspiral}

\def\mrd{\mathrm{d} }
\def\pd{\partial}
\def\m{\tx{m}}
\def\kg{\tx{kg}}
\def\s{\tx{s}}
\def\rsp{r_\tx{sp}}
\def\d{\Delta }
\def\dt{\delta}
\def\hz{\text{Hz}}
\def\be{\begin{equation}}
\def\ee{\end{equation}}
\def\ba{\begin{eqnarray}}
\def\ea{\end{eqnarray}}
\def\nn{\nonumber}
\newcommand{\tx}[1]{\text{#1}}
\newcommand{\scf}[1]{\times 10^{#1}}
\newcommand{\td}[1]{\tilde{#1}}
\newcommand{\mc}[1]{\mathcal{#1}}
\newcommand{\secref}[1]{Sec.~\ref{#1}}
\newcommand{\eqnref}[1]{(\ref{#1})}
\newcommand{\figref}[1]{Fig.~\ref{#1}}
\newcommand{\appref}[1]{Appendix~\ref{#1}}
\renewcommand{\eqref}[1]{Eq. (\ref{#1})}

\begin{abstract}
In the typical data analysis and waveform modeling of the \acp{GW} signals for \acp{BBH},
it's assumed to be isolate sources in the vacuum within the theory of \ac{GR}.
However, various kinds of matter may exist around the source or on the path to the detector,
and there also exist many different kinds of modified theories of gravity.
The effects of these modifications can be characterized within the \ac{ppE} framework,
and the corresponding phase corrections on the waveform at leading \ac{PN} order are also expressed by the additional parameters for these effects.
In this work, we consider the varying-G theory and the dynamical friction of the dark matter spike as an example.
Both of these two effects will modify the waveform at -4\ac{PN} order, if we choose the suitable power law index for the spike.
We choose to use a statistic to characterize the dispersion between the posterior of $\dot G$ for different events.
For different astronomical models, we find that this statistic can distinguish these two models very effectively.
This result indicates that we could use this statistic to distinguish other degenerate effects with the detection of multiple sources.
\end{abstract}

\maketitle

\section{Introduction}

In the 2030s there expected to have multiple space-borne \ac{GW} detectors on operation,
including TianQin \cite{Luo:2015ght,TianQin:2020hid,Luo:2025sos}, LISA \cite{Danzmann:1997hm, LISA:2017pwj}, and Taiji\cite{Hu:2017mde}.
These detectors will focus on the millihertz frequency band of \ac{GW},
and observe many different kinds of sources, such as Galactic Compact Binaries \cite{Korol:2017qcx, Huang:2020rjf},
\acp{MBHB} \cite{Klein:2015hvg,Wang:2019ryf,Feng:2019wgq},
the inspiral of Stellar Mass Black Hole Binaries \cite{Sesana:2016ljz, Kyutoku:2016ppx,Liu:2020eko}, \acp{EMRI} \cite{Babak:2017tow,Fan:2020zhy},
and Stochastic \ac{GW} Background \cite{Caprini:2015zlo, Bartolo:2016ami, Liang:2021bde, Cheng:2022vct}.
Due to the longer duration and larger \ac{SNR} of the signals,
the \ac{PE} precision will also be much higher.
Thus it will have significant improvements on the study of fundamental physics \cite{Luo:2025ewp,Gair:2012nm, Barausse:2020rsu, LISA:2022kgy,Perkins:2020tra}, astrophysics \cite{Li:2024rnk,LISA:2022yao, Baker:2019pnp} and cosmology \cite{Tamanini:2016zlh, Zhu:2021aat,LISACosmologyWorkingGroup:2022jok, Caldwell:2019vru,Zhu:2021aat,Zhu:2021bpp}.
One of the most important topics is to study the nature of gravity and \acp{BH} in the strong field regime \cite{Shi:2019hqa,Shi:2022qno,Zi:2021pdp,Xie:2022wkx,Kong:2024ssa,Tan:2024utr,Rahman:2022fay}.

However, most of these analyses have assumed ideal situations.
According to the analysis in \cite{Gupta:2024gun,Dhani:2024jja,Lau:2024idc,Garg:2024qxq,Chandramouli:2024vhw}
, many different effects may cause false \ac{GR} violation in the observation of \ac{GR}.
These effects can be classified into noise systematics, waveform systematics, and astrophysical aspects.
Then it will be very important to deduce the systematics and distinguish the origin of the possible deviation.

In this work, we will consider the possibility of environmental effects.
Although most of the analysis assumed the \ac{GW} generated by isolated sources propagating in the vacuum, there could exist matter surrounding the source or laying on the path.
For the matter laying on the path, gravitational lensing will happen\cite{Lin:2023ccz,Tambalo:2022wlm,Caliskan:2022hbu}.
For the matter surrounding the source, there could be dark matter halo, the accretion disk, and the third body\cite{Camilloni:2023xvf,Tahelyani:2024cvk,AbhishekChowdhuri:2023rfv}.
For the binaries immersed in these environments, they will experience accretion, gravitational pull and dynamical friction\cite{Barausse:2014tra}.
In this work, we will focus on the dark matter spike \cite{dmagc}, which is predicted to exist around the \acp{MBHB} in the center of the galaxies.
The existence of dark matter spike has not been confirmed by current observations, and the future \ac{GW} detectors are expected to verify whether it exist or not.
It's pointed out in\cite{Barausse:2014tra} that the dynamical friction will dominate the evolution of the binaries, and thus it should be the leading order effect.
The waveform corrections of dark matter mini-spikes, through gravitational pull and dynamical friction force are investigated in \cite{eda,pmm}.
The eccentricity is also taken into account in \cite{gyy}.
With the observed events of LIGO, the density is constrained in \cite{fclg}.

The detectability of \ac{DM} spike with \acp{IMRI} and \acp{EMRI} for space borne detectors are analyzed in \cite{Cole:2022yzw}.
They also discussed the distinguishability of \ac{DM} spike and the accretion disk and gravitational atom.
The Bayesian analysis of \ac{DM} spike around an \acp{IMRI} system for LISA is studied in \cite{mdme}.
The possible bias on the parameter estimation for LISA is analyzed in \cite{pdppe}, both the misuse of vacuum template and the uncertainty of the ppE template will cause significant bias.

Although the best source to constrain the environmental effects is the binaries with large mass ratios\cite{Zhao:2024bpp,Rahman:2023sof},
but the expected detection number of the extreme mass-ratio inspirals in some pessimistic models is almost zero.
So we will focus on \acp{MBHB} in this work, since we could detect a few events even in the most pessimistic model.
The phase correction due to the environment effect can be expressed in the \ac{ppE} formalism \cite{Yunes:2009ke,ppewv},
which is a parameterized framework to describe the deviation from \ac{GR} on \acp{GW}.
For the dynamical friction of dark matter spike with a power-law density distribution for which the power index is $\gamma=\frac{3}{2}$ and the varying-G theory,
the phase corrections are both introduced at $-4$ PN order.
Thus the existence of dark matter may cause false positive result on the varying of gravitational constant.
Although the chosen value of $\gamma$ exceeds the physical range given by \cite{dmagc}, but the degenerate with other modified theories of gravity at a different PN order may also happen.

However, the varying rate of the gravitational constant should be a constant for all the sources, while the property of the dark matter surrounding each source could be different.
With a single event, it will be difficult to distinguish these two effects.
But with multiple events, if the deviation is caused by the varying of $G$,
we will get a consistent result on the posteriors of $\dot G$.
On the other hand, if the deviation is caused by \ac{DF} and we misinterpret it as the varying-G theory,
we will get a disperse result on the posteriors.
Thus by analyzing the dispersion of the posteriors, we can distinguish these two effects.

This paper is organized as follows.
In Section \ref{sec:model}, we give a brief review on the waveform and its correction within the ppE framework, then we focus on the \ac{DF} of \ac{DM} spike for environment effect, and the varying $G$ theory for modified gravity.
Then, in Section \ref{sec:dm}, we introduce the \ac{FIM} used for \ac{PE}, and analysed the precision on the measurement of the characteristic density of the \ac{DM} spike $\rho_0$ for both varying the source parameters and using the astronomical models.
In Section \ref{sec:stat}, we give the correspondence between $\rho_0$ and $\dot G$, and introduce the statistic to characterise the dispersion of the posterior for the events observed.
Then we calculate the distribution of the statistic for three different models, and analyzed the distinguishability between different models.
We give a brief summary and discussion in Section \ref{sec:con}.
Throughout this work, the geometrized unit system ($G=c=1$) is used.

\section{Waveform Model}\label{sec:model}

There exist many different kinds of models for the modified theory of gravity and the environment surrounding the binaries.
However, all the effects on the waveform can be characterised within the \ac{ppE} framework.
In this section, we first introduce the \ac{ppE} formalism with higher mode phase corrections.
Then for a specific model, we introduce the waveform correction due to the dynamical friction of the dark matter spike.
We also introduce the waveform correction of the varying-G theory, which will be degenerate with dynamical friction if we choose $\gamma=\frac{3}{2}$

\subsection{the ppE framework}

In order to describe the leading order corrections on the waveform in a parametric form, the \ac{ppE} formalism was introduced \cite{bppE}.
When the masses of the components are highly asymmetric, higher modes of gravitational waves will be important in the waveform.
According to \cite{gjmfk}, the \ac{ppE} formalism for waveforms with higher modes is
\begin{equation}
\td{h}(f)=\sum_{lm}\td{h}_{lm}(f)=\sum\td{h}^\text{GR}_{lm}(f)e^{i\delta\Psi_{lm}}.
\end{equation}
In this work, we use IMRPhenomXHM \cite{arXiv:2001.10914} to obtain the \ac{GR} waveform with higher modes.
Since the amplitude correction is nearly negligible, in this work, we will only consider the phase correction which is,
\begin{equation}
\delta\Psi_{lm}=\beta_{lm}u^{b_{lm}}.
\end{equation}
$u=(\pi\mc{M}_cf)^{1/3}$ represents the relative velocity of the binary components, and $\mc{M}_c=(m_1m_2)^{3/5}/(m_1+m_2)^{1/5}$ is the chirp mass for the binary with component masses $m_1$ and $m_2$.
We will also use the total mass $M=m_1+m_2$ and the symmetric mass-ratio $\eta=m_1m_2/(m_1+m_2)^2$ in the following parts.
The corrections for different higher modes can be represented by that of $22$ mode as
\begin{equation}
\beta_{lm}=\left(\frac{2}{m}\right)^{\frac{b_{lm}}{3}-1}\beta_{22},~~~b_{lm}=b_{22}.
\end{equation}
In the following part, we will use $\delta\Psi$ to represent $\delta\Psi_{22}$ for simplicity.

\subsection{Dynamical Friction of Dark Matter Spike}

If the dark matter is composed of massive particles,
then due to the coevolution of the central \ac{BH} and the dark matter halo,
the adiabatic growth of the \ac{BH} will give rise to a spike \cite{dmagc}.
Then the density of the dark matter in the spike is much higher than the average density of dark matter in the surrounding halo.
The density profile of the spike often has a steeper slope and follows a power law relationship close to the central object.
\begin{equation}
\rho(r)=\rho_0\left(\frac{r_0}{r}\right)^\gamma
\end{equation}
where $r$ is the distance to the center of mass of the binary.
According to the analysis in \cite{dmagc}, the physical range of the power index $\gamma$ is $[2.25,2.5]$.
We have neglected the inner radial cutoff of the spike, since it will be smaller than the radius of the \ac{ISCO} for the binary.
Then the total mass enclosed in the sphere with radius $r$ is
\begin{equation}
m(r)=\frac{4\pi\rho_0r_0^\gamma}{3-\gamma}r^{3-\gamma}\label{mass}
\end{equation}

Besides the characteristic radius $r_0$, where the density is equal to $\rho_0$,
we can also define $r_{200}$ as the radius where the density is $200$ times the critical density of the matter $\rho_{cm}$, which is determined by the critical density of our universe $\rho_c=9.73\times 10^{-27}\text{kg}/\text{m}^3$ \cite{WMAP1OB} and the redshift $z$.
Thus we will have
\begin{equation}
\rho(r_{200})=\rho_0\left(\frac{r_0}{r_{200}}\right)^\gamma=200\rho_{cm}
\end{equation}
If $M$ is the total mass of the central bianry \ac{BH}, we can determine the parameters of the spike by assuming that
\begin{eqnarray}
m(5r_0)&=&{4\pi\rho_0r_0^\gamma\over 3-\gamma}(5r_0)^{3-\gamma}=2M\\
m(r_{200})&=&\frac{4\pi\rho_0r_0^\gamma}{3-\gamma}r_{200}^{3-\gamma}=NM.
\end{eqnarray}
By solving these equations, we will have
\begin{eqnarray}
\rho_0&=&200 \rho_{cm}5^\gamma\left(\frac{N}{2}\right)^\frac{\gamma}{3-\gamma}\label{equ:rho0}\\
r_0&=&\left(\frac{M(3-\gamma)}{2\pi\rho_0}\right)^\frac{1}{3}5^\frac{\gamma-3}{3}.\label{equ:r0}
\end{eqnarray}
In previous studies \cite{eda,pmm,mdme,galdy,Sesana_2014} $N$ is assumed to be in the range of  $10^3-10^6$.
Although $N$ will not be same for all the spikes in our universe, its dispersion will cause to the dispersion on the value of $\rho_0$.
Then the dispersion on the corresponding parameter of the degenerate modify gravity effects will be much larger.
So, in this study, we will fix $N$ to be $10^6$, then if we choose a special value of $\gamma$, both $\rho_0$ and $r_0$ are determined according to the above formulas.

For the binary immersed in the spike, due to the varying complexion of the near dark matter, the fluctuating force acting on each \ac{BH} will form the dynamical friction.
Thus the \acp{BH} will have a systematic tendency to be decelerated in the direction of its motion.
For an object with mass $m$ moving in the spike with the orbital radius $r$ with velocity $v$, the magnitude of the dynamical friction is
\begin{equation}
F_{DF}=\frac{4\pi\rho(r)m^2I}{v^2}
\end{equation}
$I$ is a order 1 factor depending on the velocities of components with respect to the medium\cite{kim2}.
Following the analysis in \cite{Eda:2014kra}, we use $I\cong3$ in this work.

Therefore, besides the energy loss due to the gravitational radiation, the binaries will also lose energy due to the dynamic friction.
Thus the equation for the conservation of energy is
\begin{equation}
-\frac{dE}{dt}=\frac{dE_{GW}}{dt}+\frac{dE_{DF}}{dt}\label{nlph}
\end{equation}
and we also have
\begin{equation}
\frac{dE_{DF}}{dt}=\sum_iF_{DF,i}v^i=\sum_i\frac{4\pi I\rho(r_i)m_i^2}{v_i}
\end{equation}
Here the index $i=1,2$ denotes the two \acp{BH} of the binary.
Then following the procedure in \cite{sbdrvg}, we can find that the phase correction is (similar results can also be found in \cite{canenv,dete}
\begin{equation}
\delta\Psi_\text{DF}=-\frac{75I\pi\rho_0r_0^\gamma}
{256(\gamma-8)(2\gamma-19)}Ku^{2\gamma-16},
\end{equation}
and the factor $K$ is
\begin{eqnarray}
K&=&(m_1^2m_2^{-\gamma-1}+m_2^2m_1^{-\gamma-1})\mc{M}_c\eta^{-\frac{2\gamma+2}{5}}\nn\\
&=&\frac{
\left(1+\sqrt{1-4\eta}\right)^{3+\gamma}+\left(1-\sqrt{1-4\eta}\right)^{3+\gamma}}
{\mc{M}_c^{\gamma-2}\eta^{\frac{4\gamma}{5}+2}2^{\gamma+3}}\label{equ:phase}
\end{eqnarray}
so the corresponding ppE phase parameters of 22 mode are
\begin{eqnarray}
    \beta_{22}&=&-\frac{75I\pi\rho_0r_0^\gamma}
{256(\gamma-8)(2\gamma-19)}K,\\
b_{22}&=&2\gamma-16.
\end{eqnarray}
In the following analysis, we will consider $\rho_0$ as a parameter that needs to be measured, and $\gamma$ is fixed.
So $r_0$ is determined by $M$ and $\rho_0$ according to \eqref{equ:r0},
and the factor $\rho_0r_0^\gamma$ can be reduced to
\begin{equation}
\rho_0r_0^\gamma=(2\pi)^{-\frac{\gamma}{3}}5^\frac{\gamma(\gamma-3)}{3}
(3-\gamma)^\frac{\gamma}{3}\mc{M}_c^\frac{\gamma}{3}
\eta^{-\frac{\gamma}{5}}\rho_0^{1-\frac{\gamma}{3}}
\end{equation}

When $\gamma=\frac{3}{2}$, we will have
\begin{equation}
\delta\Psi_\text{DF}=-\frac{15\sqrt[4]{5}}{851968}\sqrt{\frac{3\pi\rho_0}{2}}
I\mc{M}_c\eta^{-\frac{7}{2}}H(\eta)u^{-13}
\end{equation}
and $H(\eta)$ is defined to be
\begin{equation}
H(\eta)=\left(\frac{1}{2}+\frac{\sqrt{1-4\eta}}{2}\right)^\frac{9}{2}
+\left(\frac{1}{2}-\frac{\sqrt{1-4\eta}}{2}\right)^\frac{9}{2}
\end{equation}
this means that $b=-13$ in this case, and the leading order correction is $-4$ PN.
It should be noted that, although the chose value of $\gamma$ exceeds the range of $[2.25,1.5]$, but $\gamma=\frac{3}{2}$ is also considered in previous works such as \cite{Eda:2014kra}.

\subsection{The theory of varying G}

In \ac{GR}, gravitational coupling strength $G$ is a constant independent of spacetime.
While in the varying $G$ theory,  it will change over time \cite{gdxw}.
The rate of changing $\dot{G}$ causes a phase correction \cite{ppewv} as

\begin{equation}
\delta\Psi_{\dot{G}}=-\frac{25\mathcal{S}}{851968}\dot{G}\mc{M}_c(\pi\mc{M}_cf)^{-{13\over 3}}.\label{sbyl}
\end{equation}
so the ppE parameters for the phase of 22 mode are
\begin{eqnarray}
    \beta_{22}&=&-\frac{25\mathcal{S}}{851968}\dot{G}\mc{M}_c,\\
    b_{22}&=&-13.
\end{eqnarray}

According to the analysis in \cite{gdxw,pdppe}, for general binaries, the factor $\mathcal{S}$ is
\begin{equation}
\mathcal{S}=11-\frac{35}{2}(s_1+s_2)+\frac{41}{2}\sqrt{1-4\eta}(s_1-s_2),
\end{equation}
and $s_{1,2}$ are the sensitivities for the components.
However, for \acp{BH}, we will have $s_1=s_2=\frac{1}{2}$, and thus we will have $\mathcal{S}=-\frac{13}{2}$
\cite{sbdrvg}.

We can find that the leading order correction of varying-G theory is -4 PN,
the same order as the \ac{DF} with $\gamma=\frac{3}{2}$.
Thus these two effects will have degeneracy, and if we find the -4 PN deviation in the future observations,
it will be difficult to identify the origion of this deviation.

\section{Probe the density of dark matter}\label{sec:dm}

In this section, we will analyze the capability to probe the density of the dark matter.

\subsection{fisher information matrix}

The sky-averaged noise \ac{PSD} for TianQin \cite{Wang:2019ryf} is
\begin{eqnarray}
S_n(f)&=&\frac{10}{3}\frac{1}{L^2}\left[1+\left(\frac{2fL_0}{0.41c}\right)^2\right]\nn\\
&\times&\left[\frac{4S_a}{(2\pi f)^4}\left(1+\frac{10^{-4}\hz}{ f}\right)+S_x\right]
\end{eqnarray}
where $L=\sqrt{3}\scf{5}\text{km}$ is the arm length of TianQin, $\sqrt{S_a}=10^{-15}\m\dot\s^{-2}\hz^{-1/2}$ and $\sqrt{S_x}=10^{-12}\m\dot\hz^{-1/2}$ are the acceleration and position noise of TianQin, respectively.

For signals with large \ac{SNR}, the posterior distribution of the relevant parameters can be approximated by a Gaussian distribution around the true value.
The corresponding covariance matrix is given by the inverse of Fisher matrix: $\Sigma_{ij}=(\Gamma^{-1})_{ij}$, with
\begin{equation}
\Gamma_{ij}=2\int_{f_\text{min}}^{f_\text{max}}\frac{\partial_ih(f)\partial_jh^\ast(f)+\partial_ih^\ast(f)\partial_jh(f)}{S_n(f)}df
\end{equation}
where the partial derivative $\partial_i$ corresponding to the $i$-th parameter of $\hat\theta=(\mc{M}_c,\eta,\chi_1,\chi_2,\rho_0,D_L)$
The polarization and inclination angle were not take into account since we have also averaged the orientation of the \acp{MBHB}.
We considered the ``3 month on + 3 month off'' observation scheme in our calculation, and assume all the events merged during the observation period.
Since the \ac{ppE} correction is used for the inspiral stage, we will only use the inspiral signal to do the analysis, and thus we will have
\begin{equation}
f_\text{max}=f_\text{ISCO}=\frac{1}{6\sqrt{6}\pi M}
\end{equation}
and
\begin{equation}
f_\text{min}=\left(\frac{5}{256}\right)^{3/8}\frac{1}{\pi}\mc{M}_c^{-5/8}T^{-3/8}
\end{equation}

\subsection{capability to probe the density}

For the parameters of the sources, we choose $\chi_1=0.9,\chi_2=0.8,T=5\text{yr},\phi_c={\pi\over 4}$,
and the distance is chosen to make the SNR of the events to be $100$.
$\mc{M}_c$ varies between $10^4\sim10^7M_\odot$, and $\eta$ varies between $0\sim 0.25$

\begin{figure}
	\begin{center}
		\includegraphics[scale=0.55]{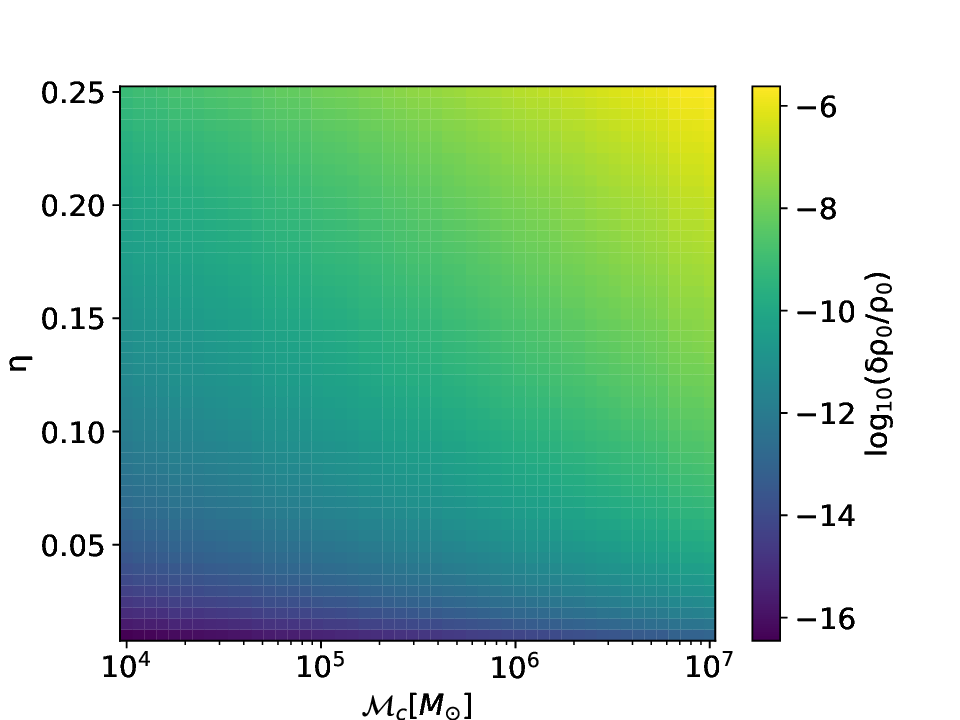}
		\caption{The capability on the \ac{PE} precision of $\rho_0$ for $\gamma=\frac{7}{3}$ and $N=10^3$ for the sources with different $\mc{M}_c$ and $\eta$.}\label{derhem2d33}
	\end{center}
\end{figure}

For the dark matter spike, a typical value for the power index is
$\gamma={7\over 3}$ according to the study of \cite{Gondolo:1999ef}.
Thus if we choose $N=10^3$, we will have
$\rho_0=8.34\times10^{-14}\text{kg}/\text{m}^3$.
The relative \ac{PE} precision for $\rho_0$ for \acp{MBHB} with different $\mc{M}_c$ and $\eta$ is plotted in \figref{derhem2d33}.
We can find that the result is much more better than the result in \cite{Cole:2022yzw,mdme}.
This is caused by the strong degeneracy between $\gamma$ and $\rho_0$.
In our Fisher analysis, $\gamma$ is fixed, then the \ac{PE} precision will be better.
Similar result for fixed $\gamma$ can also be found in \cite{Cardoso:2019rou}.

\begin{figure}
	\begin{center}
		\includegraphics[scale=0.55]{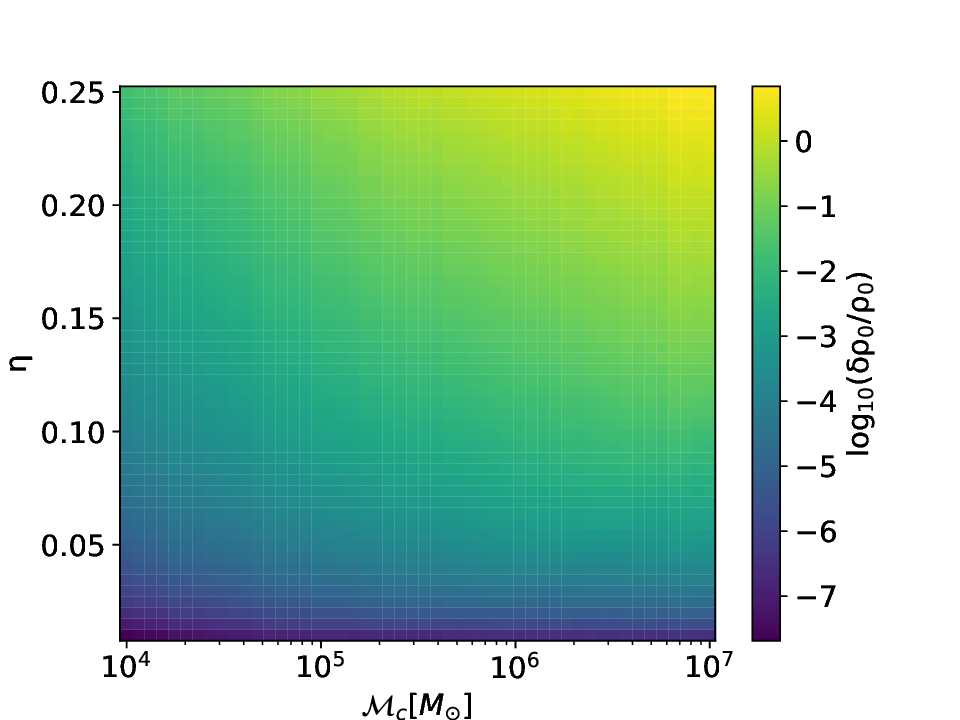}
		\caption{The capability on the \ac{PE} precision of $\rho_0$ for $\gamma=\frac{3}{2}$ and $N=10^6$ for the sources with different $\mc{M}_c$ and $\eta$.}\label{derhem1d5}
	\end{center}
\end{figure}

Since we are trying to distinguish the dark matter spike with the varying-G theory, we also considered the case for $\gamma={3\over 2}$ in \figref{derhem1d5} where we adopt $N=10^6$ and $\rho_0=4.14\times10^{-18}\text{kg}/\text{m}^3$.
$\delta\rho_0/\rho_0$ for \acp{MBHB} with different $\mc{M}_c$ and $\eta$ is plotted in \figref{derhem1d5}.

We can find that the relative precision on the measurement of $\rho_0$ could reach to the level of $10^{-18}$ for $\gamma=\frac{7}{3}$, and $10^{-7}$ for $\gamma=\frac{3}{2}$.
The result will be better for smaller $\eta$ and smaller $\mc{M}_c$.

\begin{figure}
	\begin{center}
		\includegraphics[scale=0.55]{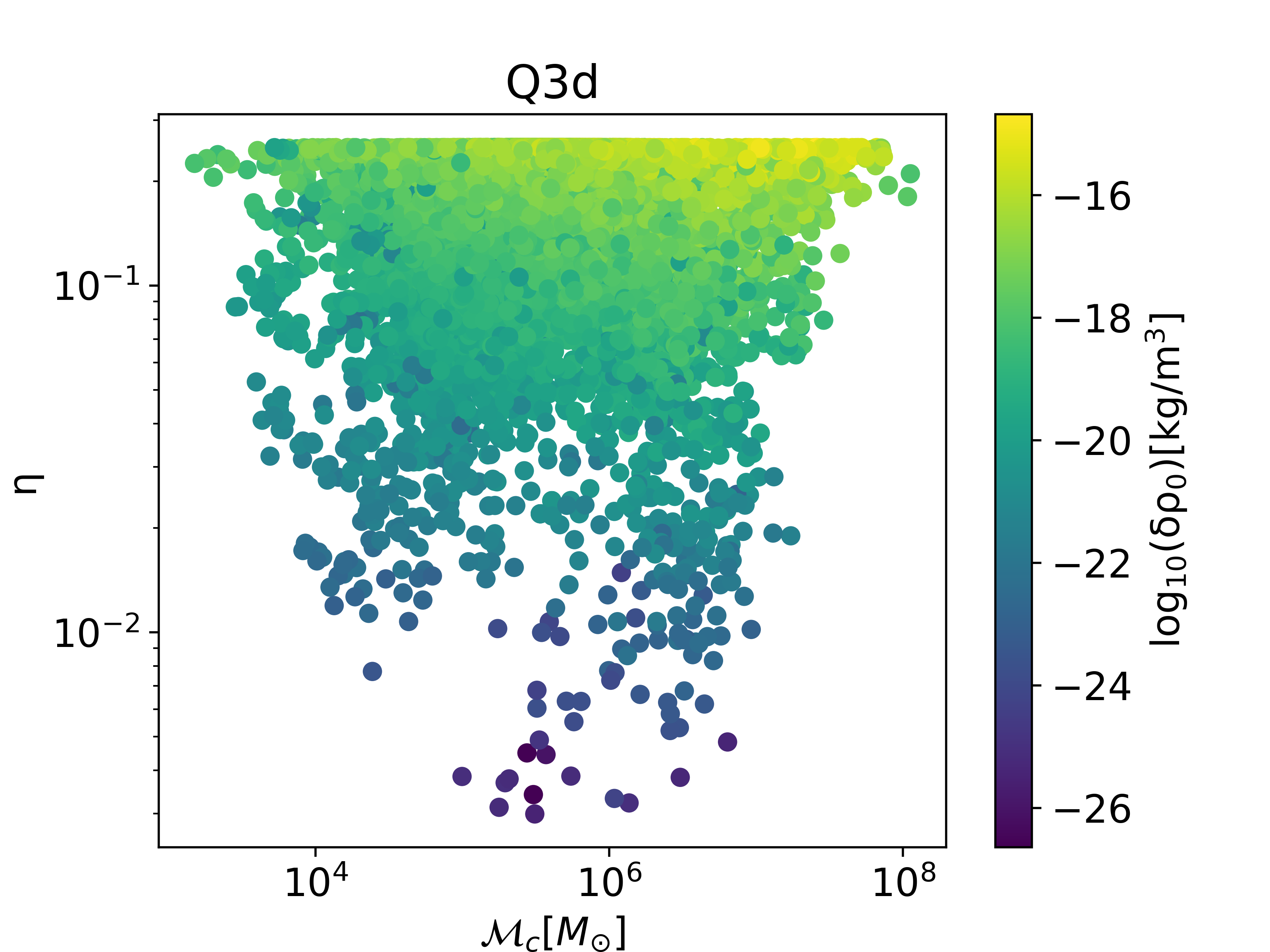}
		\caption{The precision to measure $\rho_0$ for the sources in the catalog of ``Q3\underline{~}d'' model.}\label{emq3d}
	\end{center}
\end{figure}

\begin{figure}
	\begin{center}
		\includegraphics[scale=0.55]{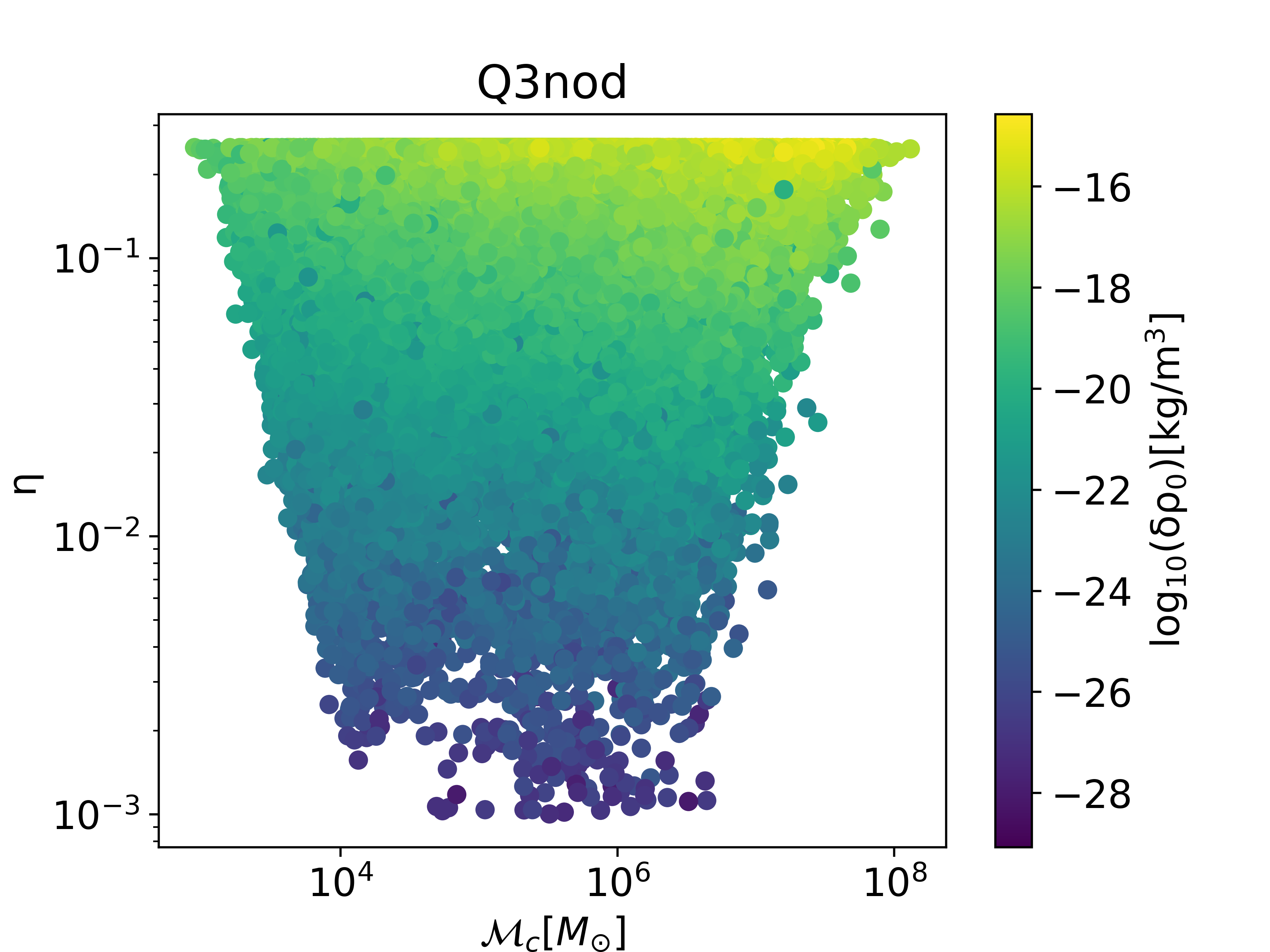}
		\caption{The precision to measure$\rho_0$ for the sources in the catalog of ``Q3\underline{~}nod'' model.}\label{emq3nod}
	\end{center}
\end{figure}

\begin{figure}
	\begin{center}
		\includegraphics[scale=0.55]{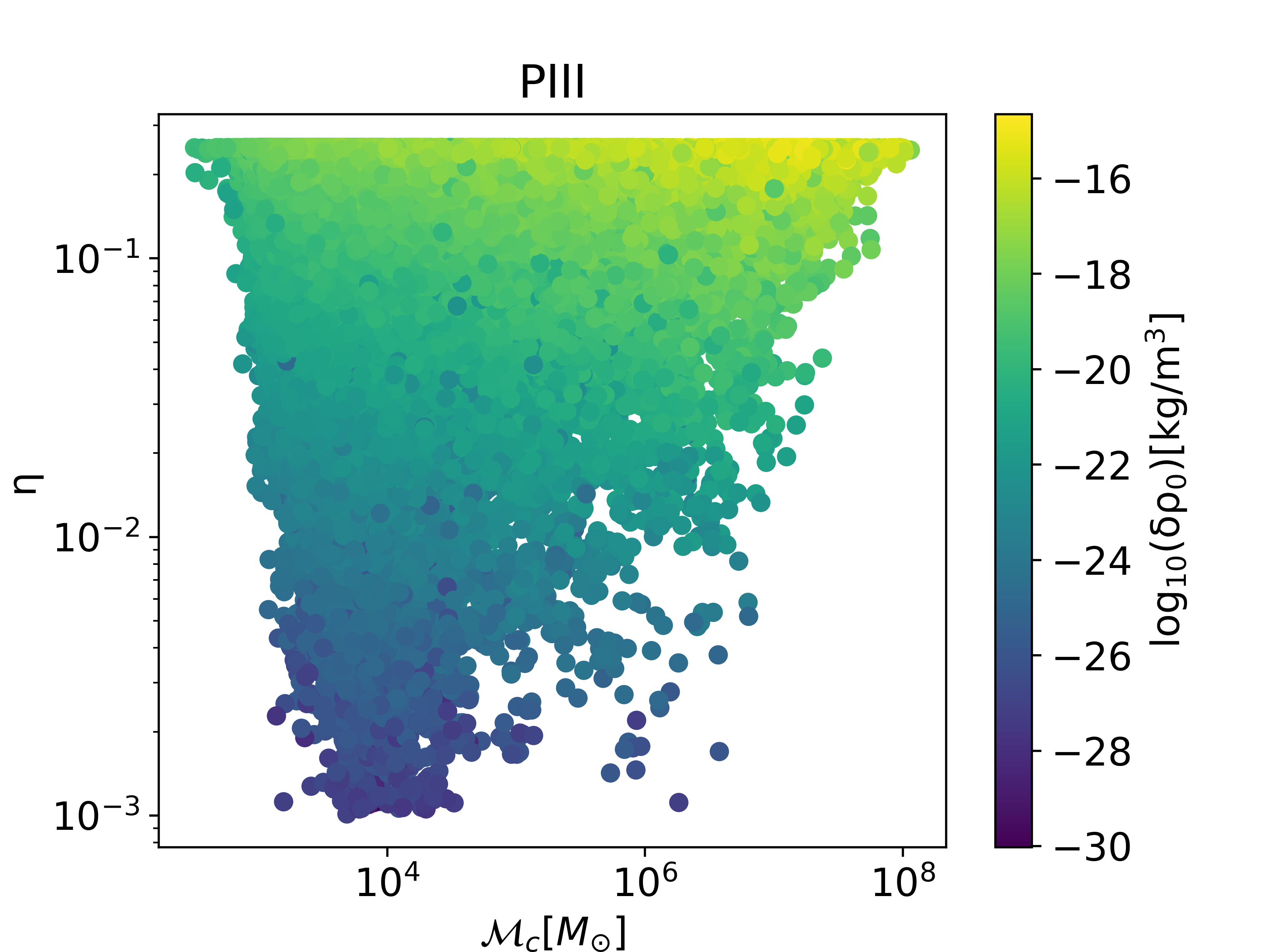}
		\caption{The precision to measure $\rho_0$ for the sources in the catalog of ``PIII'' model.}\label{emp3}
	\end{center}
\end{figure}

\begin{figure}
	\begin{center}
		\includegraphics[scale=0.55]{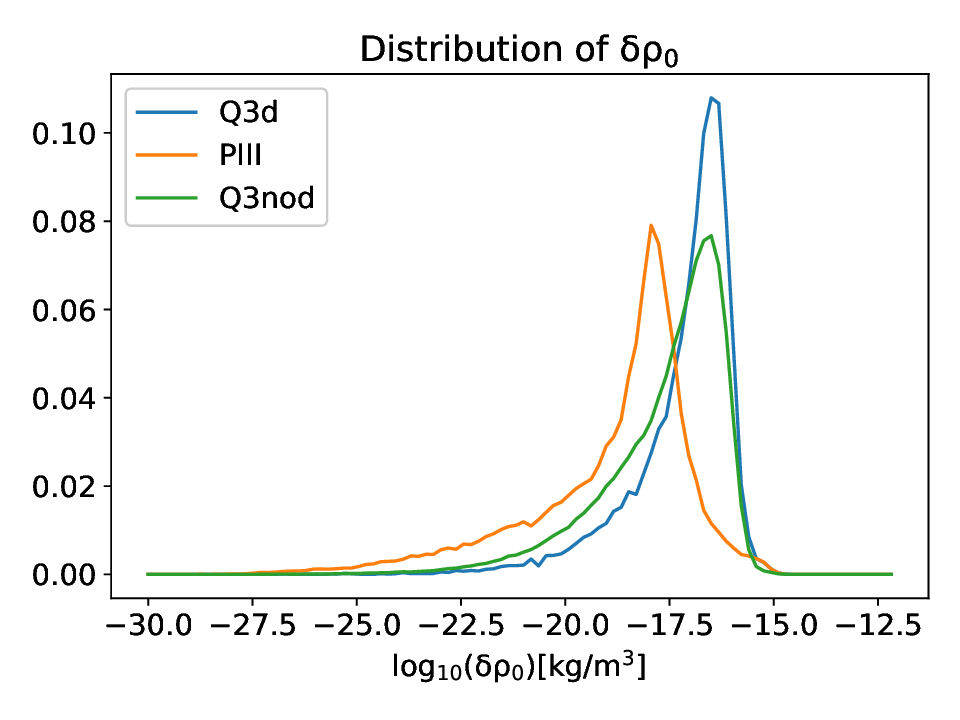}
		\caption{Distributions of $\dt\rho_0$ for ``Q3\underline{~}d'' , ``Q3\underline{~}nod'', and``popIII''}\label{drhmxdb}
	\end{center}
\end{figure}

Since we will use all the events in the observation to verify if there exist violation of \ac{GR}, we need to consider astronomical models to simulate the observation data.
We consider three different models \cite{Barausse:2012fy,Sesana:2014bea,Antonini:2015sza,Klein:2015hvg}in our work,
two heavy-seed models ``Q3\underline{~}d'' and ``Q3\underline{~}nod'', and a light-seed model ``popIII''. Each model has 1000 mock catalogues for a five-year observation with TianQin\cite{Wang:2019ryf},
and the total events number are 18112, 271444, and 56618, respectively.
We only consider the events with \ac{SNR} larger than $8$.
According to these catalogues, we can get $(z,m_1,m_2,\chi_1,\chi_2,\iota)$ of the sources, then we can calculate $\delta\rho_0$ for each source.
We use $\gamma=\frac{3}{2}, N=10^6$ in this calculation, and the results are shown in
\figref{emq3d}, \figref{emq3nod}, and \figref{emp3}.
We labeled each events with the $\mc{M}_c$ and $\eta$, and the \ac{PE} precision is shown in different colors.
We also plot the distribution of $\delta\rho_0$ in \figref{drhmxdb}.
The popIII model is shown to have a better precision on probing the density of the spike, and this is due to the fact that the events in this model may have smaller $\eta$.

We can find that for each model, there will always exist many events with the \ac{PE} precision for $\delta \rho_0$ is less than $4.14\times10^{-18}\text{kg}/\text{m}^3$.
This result implies that we can confirm the existence of  \ac{DM} spike with the observation of \acp{MBHB}.
On the other hand, if we don't consider the possibility of the existence of \ac{DM} spike, the observable correction on the phase of the signal may cause biased result when testing \ac{GR}.

\section{Distinguishing dark matter DF and $\dot{G}$ effect}\label{sec:stat}

Since both the dynamical friction for the dark matter spike with $\gamma=\frac{3}{2}$, and the varying-G theory will modify the waveform at $-4$ PN order,
if we find the deviation at this order in future detections,
it will be a very important question to verify which theory is correct.
If we neglect the environment effect, we may get false positive results in testing \ac{GR}.
It will be difficult to distinguish these two effects through data analysis of a single event.
However, if the detected deviation is caused by varying-G, then the corresponding $\dot G$ will be the same for all the events.
But if it's caused by \ac{DF}, the corresponding $\dot G$ will be different for each events.
In this section, we will use the results of \ac{PE} for all the events in the observation period to distinguish these two effects.

\subsection{The relation between DF and $\dot{G}$ theory}

If we set the phase correction of \ac{DF} equal to that of varying-G,
we can get that
\begin{eqnarray}\label{equ:relation}
\dot{G}=-\frac{3\sqrt[4]{5}}{65}\sqrt{6\pi\rho_0}I\eta^{-\frac{7}{2}}H(\eta)
\end{eqnarray}
According to \eqref{equ:rho0}, $\rho_0$ only dependent on $\rho_{cm}$, $\gamma$, and $N$, which are all fixed value in our calculation.
Then it should be a constant for all the events in our analysis.
However, since $\eta$ will be different for each event,
the corresponding $\dot G$ will also be different.
It should also be noticed that the corresponding $\dot{G}$ is always negative
according to \eqref{equ:relation}

Then for a \acp{MBHB} in a dark matter spike with $\gamma=\frac{3}{2}$,
we can pretend it's caused by the varying of gravitational constant.
Then using the value of $\dot G$ obtained with \eqref{equ:relation},
we can get the \ac{PE} precision of each event.

\subsection{The statistic $F$ to distinguish DF and $\dot{G}$}

If the modification is caused by varying-G theory,
then we should obtain the same value of $\dot G$ for all the events.
But if the modification is caused by the dynamical friction,
according to \eqref{equ:relation}, the corresponding $\dot G$ for each event will be different.
So by checking if all the events have a consistent prediction of $\dot G$,
we can distinguish these two effects.
Thus we can use a statistic to describe the consistency of the values of $\dot G$ given by all the events.

We can assume that we have detected $n$ events in our observation,
and we find deviations at $-4$ PN order.
If we make the hypothesis that the deviation is caused by $\cot G$,
then for the $i$-th event, the true value of $\dot G$ is $\dot G_i$,
and the corresponding \ac{PE} precision is $\delta\dot G_i$.
So the posterior of $\dot G$ for the $i$-th event can be approximated as
a normal distribution: $\dot G\sim N\left(\dot G_i,\delta\dot G_i\right)$.
In the real data analysis, the posterior of $\dot G$ may not be Gaussian, and we could use the mean value and the standard deviation to calculate the following statistic.
Since the result of Fisher analysis will have difference on the order of magnitude with the result of Bayesian,
the behavior of the statistic will not change too much in the analysis of the real data.
Due to the existence of instrument noise,
the center of this distribution $\dot G_i^\prime$ will deviate from the true value,
and it should also obey this distribution $\dot G_i^\prime\sim N\left(\dot G_i,\delta\dot G_i\right)$.

Then for all the events, we can define the mean value of the central value of each event as
\begin{equation}
\overline{\dot G^\prime}=\frac{1}{n}\sum_{i=1}^n\dot G_i^\prime.
\end{equation}
Then we can define a statistic $F$ to characterize the dispersion of the posteriors of all the events.
\begin{equation}
F=\frac{\sum_{i=1}^n\left(\dot G^\prime_i-\overline{\dot G^\prime}\right)^2}{\sum_{i=1}^n\delta G_i^2}\label{equn:stat}
\end{equation}

If all the events have the same central value, we will have $F=0$.
However, due to the random error in the \ac{PE}, it will be a small number instead of exactly zero.
On the other hand, if the central values are totally different,
and the \ac{PE} precision is much smaller than the differences,
we will get a very large $F$.

\begin{figure}
	\begin{center}
		\includegraphics[scale=0.55]{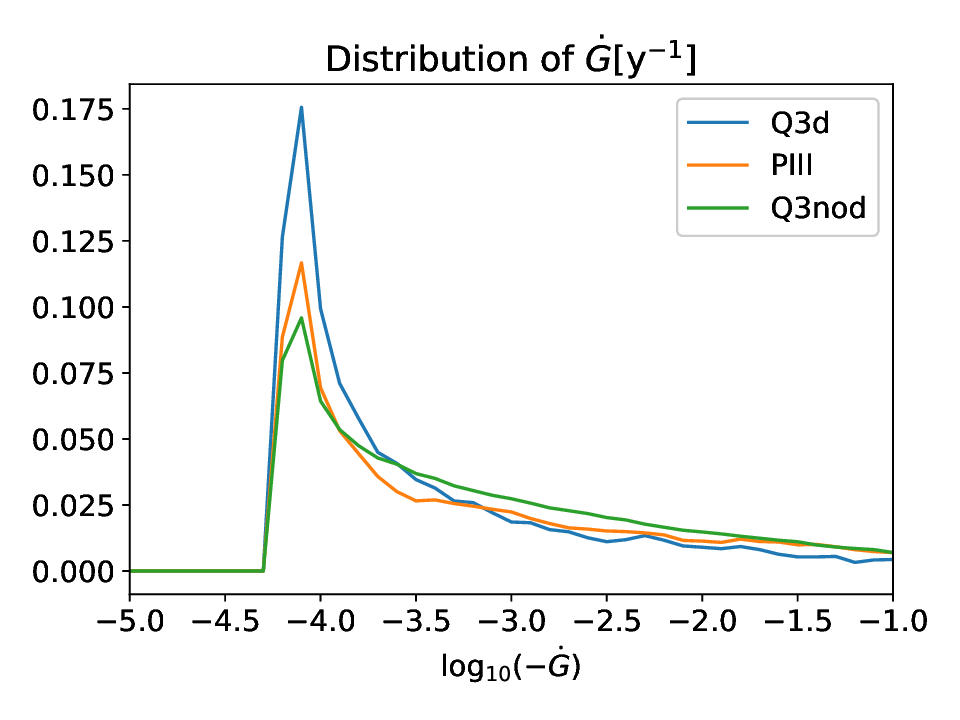}
		\caption{Distributions of $\dot{G}$ considering Q3d, PIII and Q3nod.}\label{trhgd}
	\end{center}
\end{figure}

For dynamical friction, $\dot G_i$ is obtained with \eqref{equ:relation}.
The distribution of $\dot G_i$ for each model is shown in \figref{trhgd}.
We can see that for the chosen parameters of the \ac{DM} spike and the \ac{MBHB},
the value of $\dot G_i$ is about $-10^{-4}$.
Since $\rho_0$ is a constant if we fix $N$ and $\gamma$,
the distribution of $\dot G$ is determined by the distribution of $\eta$.
The cut-off on the left part of \figref{trhgd} corresponds to the equal mass events with $\eta=1/4$.
For the case of varying-G theory, $\dot G_i$ is a constant.
In our analysis, we consider three situations, corresponding to
$\dot G=0,-10^{-14},-10^{-4} \text{yr}^-1$ respectively.
Here $0$ corresponding to \ac{GR}, and $10^{-14}$ is about the current constraint of $\dot G$, while $-10^{-4}$ is the typical value caused by dynamical friction as shown in \figref{trhgd}.

\begin{figure}
	\begin{center}
		\includegraphics[scale=0.55]{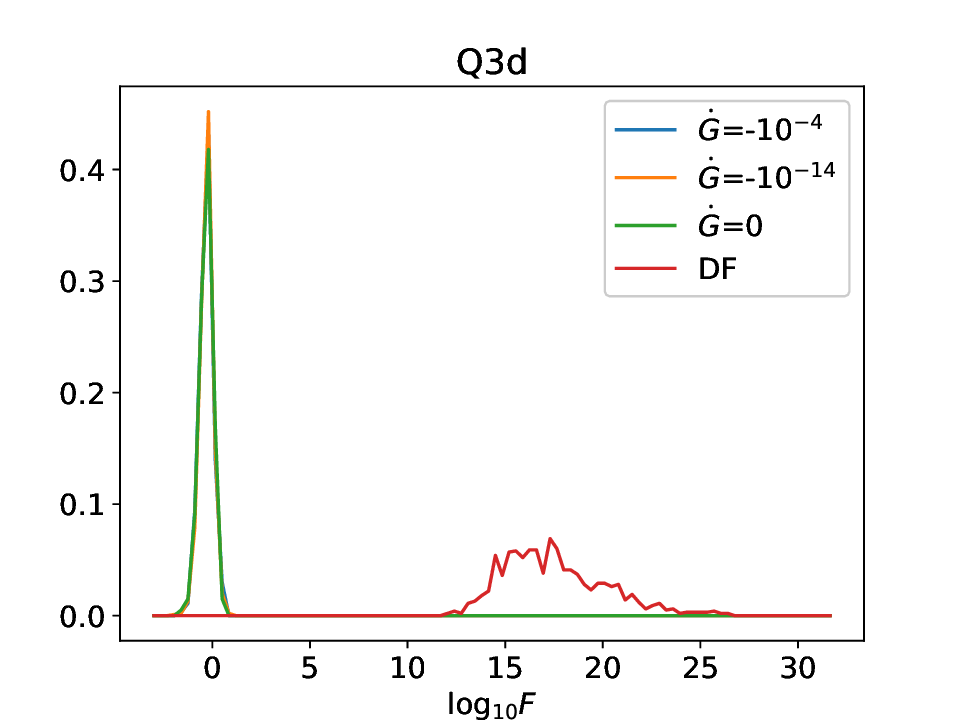}
		\caption{The distribution of $F$ for Q3\underline{~}d model for different values of $\dot G$ in varying-G theory and the dynamical friction of dark matter spike with $\gamma=\frac{3}{2}$.}\label{fjhq3d}
	\end{center}
\end{figure}

\begin{figure}
	\begin{center}
		\includegraphics[scale=0.55]{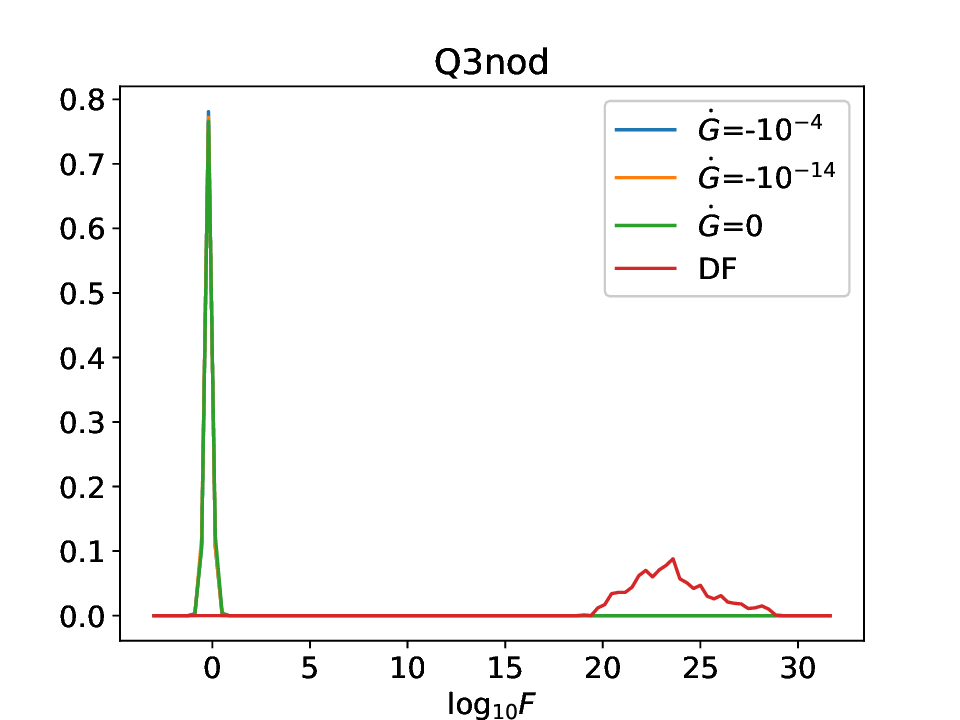}
		\caption{The distribution of $F$ for Q3\underline{~}d model for different values of $\dot G$ in varying-G theory and the dynamical friction of dark matter spike with $\gamma=\frac{3}{2}$.}\label{fjhq3nod}
	\end{center}
\end{figure}

\begin{figure}
	\begin{center}
		\includegraphics[scale=0.55]{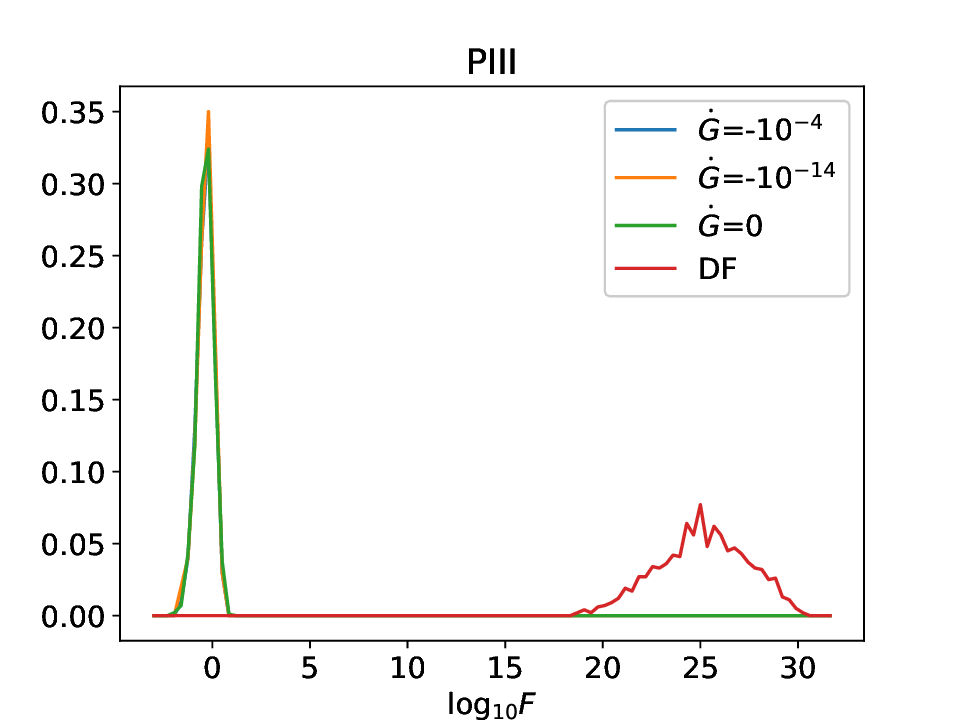}
		\caption{The distribution of $F$ for Q3\underline{~}d model for different values of $\dot G$ in varying-G theory and the dynamical friction of dark matter spike with $\gamma=\frac{3}{2}$.}\label{fjhp3}
	\end{center}
\end{figure}

For each catalog of a specific astronomical model, we can calculate the statistic $F$ as the procedure we described above.
Then we can get the distribution of $F$ with all the 1000 catalogs for each case.
The results are shown in \figref{fjhq3d}, \figref{fjhq3nod}, and \figref{fjhp3} respectively.
We can see that for all three models, the distribution of the statistic is totally different for varying-G theory and dynamical friction.
No matter what the true value of $\dot G$ is, all the three choices will give similar distributions of $F$, since the \ac{PE} precision $\delta \dot G$ is not sensitive to the value of $\dot G$.
For dynamical friction, $F$ will be larger than $10^{10}$, and it will be smaller than $10$ for varying-G theory.
Since the distribution of $F$ for these two cases are totally different without any overlap, we have a very wide range to shoose the separatrix of these two effects.
Thus in the real detection, if we get a $F$ much larger than $10$,
we can say it's caused by the environmental effect definitely.

\section{Conclusion}\label{sec:con}

In this work, we make a preliminary attempt to distinguish the effect of modified theory of gravity and of environment on \acp{GW}.
We choose the \ac{DF} of the dark matter spike with $\gamma=\frac{3}{2}$, and the varying-G theory as an example.
Both of these effects have the leading order modification at $-4$ PN order.
Since spaceborn \ac{GW} detectors are much more sensitive to the lower order PN corrections, the characteristic density $\rho_0$ can be measured to the level of $10^{-20}\text{kg}/\text{m}^3$ and relative precision of $10^{-7}$.

Then with three different models, we use a statistic $F$ to describe the dispersion of the measured $\dot G$ for the simulated catalogue.
We find that if the deviation is caused by the theory of varying-G, $F$ will be smaller than $10$.
On the other hand, if the deviation is caused by dynamical friction, $F$ will be much large, to the level of $10^10$.

However, we have made a lot of simplified assumptions in our analysis, which need to be improved in future work.
First, we have assumed $N=10^6$ and $\gamma=\frac{3}{2}$ for all the events in \ac{DF}, and thus $\rho_0$ is also fixed.
In practice, all these parameters could be different for each event.
However, if $\gamma\neq\frac{3}{2}$, the modification will introduce in other order, and will not be confused with varying-G theory.
On the other hand, if $\rho_0$ also have different values for each event,
the dispersion of the distribution of measured $\dot G$ will be larger,
and thus we will get a larger $F$.
So our method is still valid in these more complex cases.

Another problem is that the central value of $\dot G$ corresponding to \ac{DF} is much larger than current constraint.
Thus if we detected such a large result of $\dot G$, it could not be caused by the modified theory of gravity definitely.
This is true for the cases we considered in this work.
However, there exist many different kinds of environmental effects and modified theories of gravity which could have corrections on the \ac{GW} at the same order.
For example, the phase correction of a stellar-orgin \ac{BBH} caused by the peculiar acceleration as it's orbiting around a \ac{MBH} \cite{Tamanini:2019usx} and large extra dimensions \cite{Yagi:2011yu} also introduced at -4 PN order.
And according to \cite{Barausse:2014tra,Chamberlain:2017fjl},the phase correction due to the electric charge and the mass of graviton are all introduced at 1 PN order.
If we choose some specific power index for the density distribution of dark matter or accretion disk, the corresponding phase correction due to gravitational pull or dynamical friction may also be degenerate with other modified theory of gravity.
So, this method may also be used to distinguish these effects.
In fact, we may use this method to distinguish the effects at the same PN order if the dependence of their phase correction on the source parameter are different.
For example, both varying-G and large extra dimension theory introduce the correction at -4 PN, but $\beta_{ED}\propto\frac{3-26\eta+34\eta^2}{\eta^{2/5}(1-2\eta)}$, while $\beta_{\dot G}\propto\mathcal{M}_c$.
So we can distinguish these two theories with multiple events.

However, if two theories are totally degenerated with each other, which means that for different source parameters, the parameter in the first theory corresponding to a specific value of the parameter in the second theory,
Our method will be useless in the distinguishment.
For example, the correction due to the Scalar-Tensor theory and Einstein-\AE ther theory are all introduced at -1 PN order, but the corresponding $\beta$ are all proportional to $\eta^{2/5}$.
Then this method will be useless in this case.

\begin{acknowledgments}\label{sec:acknowledgments}

The authors thank Yi-Ming Hu, Changfu Shi, Xiangyu Lyu for helpful discussion.
This work is supported by the Guangdong Basic and Applied Basic Research Foundation(Grant No. 2023A1515030116), the National Key Research and~Development Program of China (Grant No. 2023YFC2206703), the Guangdong Major Project of Basic and Applied Basic Research (Grant No. 2019B030302001), and the National Science Foundation of China (Grant No. 12261131504).
\end{acknowledgments}

\bibliography{DMDFGD}

\end{document}